\documentclass[a4paper,12pt]{article}

\usepackage[pdftex]{graphicx}
\usepackage{latexsym,amsfonts,amsmath,amscd,epsfig}
\usepackage{amssymb,amsxtra,mathtools}
\usepackage{url}
\usepackage{pdflscape}
\newcommand{\bec}{\begin{center}}
\newcommand{\ec}{\end{center}}
\newcommand{\bee}{\begin{equation}}
\newcommand{\ee}{\end{equation}}

\DeclareMathAlphabet{\mathbi}{OML}{cmm}{b}{it}
\title{ The study of the properties of the
 extended Higgs boson sector within hMSSM model}

\author{T.V. Obikhod\thanks{E-mail: obikhod@kinr.kiev.ua}, E.A. Petrenko\\
\small\emph{Institute for Nuclear Research, National Academy of Science of Ukraine} \\
\small\emph{47, prosp. Nauki, Kiev, 03028, Ukraine}}

\date{\small\today}
\begin{document}
\maketitle

\abstract{ Using the latest experimental data, performed by ATLAS 
Collaboration and  within the framework of the Minimal Supersymmetric 
Standard Model, we presented the   calculations for cross sections
 times branching fractions, $\sigma\times Br$, as a functions of the
  CP-even, H, Higgs boson  mass, CP-odd, A, Higgs boson mass and charged, 
  H$^{\pm}$, Higgs boson mass. Using the restricted parameter set, received from 
  the hMSSM+HDECAY and "low-tb-high" scenarios, with the help of 
  the computer programs SOFTSUSY, Prospino and SusHi, we received 
  the large values of $\sigma\times Br$ for A and H bosons at tan$\beta$=2 
  for the planned 14 TeV at the LHC and found the large $\sigma\times Br$ 
  at tan$\beta$=30 for charged Higgs boson. The obtained results 
  are of experimental interest as they are 
  connected with the experimental searches for new physics beyond 
  the Standard Model at the LHC.}\\
\vspace*{3mm}\\
PACS: 11.25.-w, 12.60.Jv, 02.10.Ws\\




\newpage
\section{Introduction}
The searches for supersymmetry (SUSY) are motivated by the 
solutions of the most important problems: the hierarchy problem, 
gauge coupling unification and dark matter problem \cite{1.}. 
Experimental searches 
for SUSY in the most probable channels for the superparticle 
production at the LHC did not lead to the desired results and 
set new lower limits in the mass range about 2 TeV for gluino 
and squarks \cite{2.}. This fact led to the need for 
SUSY searches in other sectors, for example, in the electroweak 
sector. As highlighted in CERN Courier \cite{3.}:
"Based on data recorded in 2016, CMS has 
covered models of electroweak production of "wino"-like 
charginos and neutralinos with searches in different final states. 
More results are expected soon, and the sensitivity of the searches 
will largely profit from the extension of the data set in the 
remaining two years of LHC Run 2". Another important sector 
for SUSY searches in low mass range of 1 TeV are the searches 
for extended Higgs boson sector predicted by Minimal Supersymmetric 
Standard Model (MSSM) \cite{4.}, that consists of five Higgs bosons: 
CP even Higgs bosons, h and H, CP odd Higgs boson, A, charged 
Higgs bosons, H$^{\pm}$.

    The purpose of our paper is to calculate the production 
cross section of such particles at the energy of 14 TeV 
at the LHC in the most optimal space of parameters of the MSSM model.

\section{ Optimal parameter space for studying of the properties 
of MSSM Higgs bosons}

The masses of five Higgs bosons of MSSM model 
at tree level are calculated through the 
masses of gauge boson, M$_W$, M$_Z$, 
and two additional parameters such as  
the pseudoscalar mass, M$_A$ and  the 
ratio of vacuum expectation values of 
two Higgs doublets, tan$\beta\equiv\upsilon_u/\upsilon_d$   
\cite{5.}:
\[M^2_{H^{\pm}}=M^2_A+M^2_W\ ,\]
\[M^2_{h,H}=\frac{1}{2}\Biggl(M^2_A+M^2_Z\mp\sqrt{(M^2_A+M^2_Z)^2
-4M^2_AM^2_Z\mbox{cos}^22\beta}\Biggr)\ .\]
 
In the paper \cite{6.} the theoretical 
predictions of the MSSM Higgs particles in
 the low tan$\beta$ regime, $1 \leq$tan$\beta\leq 3$ 
 are reviewed, with the assumption that SUSY 
 should be in the range of 1 TeV. It was 
 showed that the heavier MSSM neutral H/A 
 and charged H$^{\pm}$ states can decay 
 into gauge bosons, lighter Higgs bosons 
 and top quarks, presented in Fig.1
 
\bec
{\includegraphics[width=0.85\textwidth]{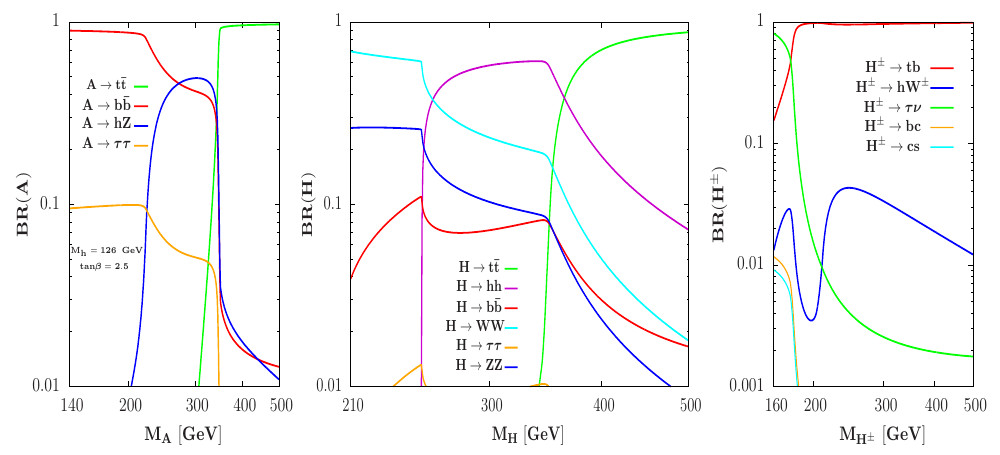}}\\
\emph{{Fig.1.}} {\emph{The branching ratios as functions of 
masses of MSSM Higgs bosons (A left, H center, H$^{\pm}$ right) 
for tan$\beta$=2.5, from \cite{6.}.}}\\
\ec

In the Handbook of LHC Higgs cross sections, 
2017 \cite{5.} are given examples of 
sensitivity on the [tan$\beta$, M$_A$] 
parameter space for the "model independent" 
hMSSM approach \cite{6.}, compared 
to the second approach \cite{7.} 
so called "low-tb-high" approach in the MSSM, 
that is orthogonal to the one previous. 
Relative differences in 
BR(H$\rightarrow$ WW) between the predictions of 
the "low-tb-high" scenario and the 
corresponding predictions obtained 
with the hMSSM+HDECAY combination are presented in Fig.2.
\bec
{\includegraphics[width=0.51\textwidth]{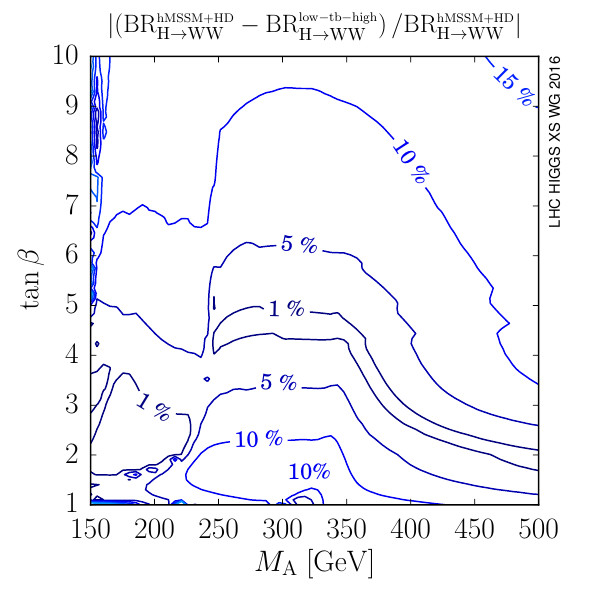}}\\
\emph{{Fig.2.}} {\emph{
Relative differences in BR(H$\rightarrow$WW) between the hMSSM+HDECAY 
scenario and the "low-tb-high" scenario, from \cite{5.}.}}\\
\ec
The results of ATLAS \cite{8.} 
and CMS \cite{9.} Collaborations 
excluded at the 95$\%$ confidence level (CL) a 
significant part of the [tan$\beta$, M$_A$] plane. 
We'll use the benchmark scenarios of the model 
independent approach for the Higgs sector, 
the hMSSM with M$_h$ = 125 GeV for the 
experimental limits on the cross sections 
times branching ratios in the context 
of the MSSM \cite{10.}. 
The results for the branching fractions 
received with the program HDECAY 
\cite{11.} for the Higgs decays in the [tan$\beta$, M$_A$]
 plane are displayed in Fig. 3 with red area 
 for the large decay rates and blue area for the small one.

\bec
{\includegraphics[width=0.75\textwidth]{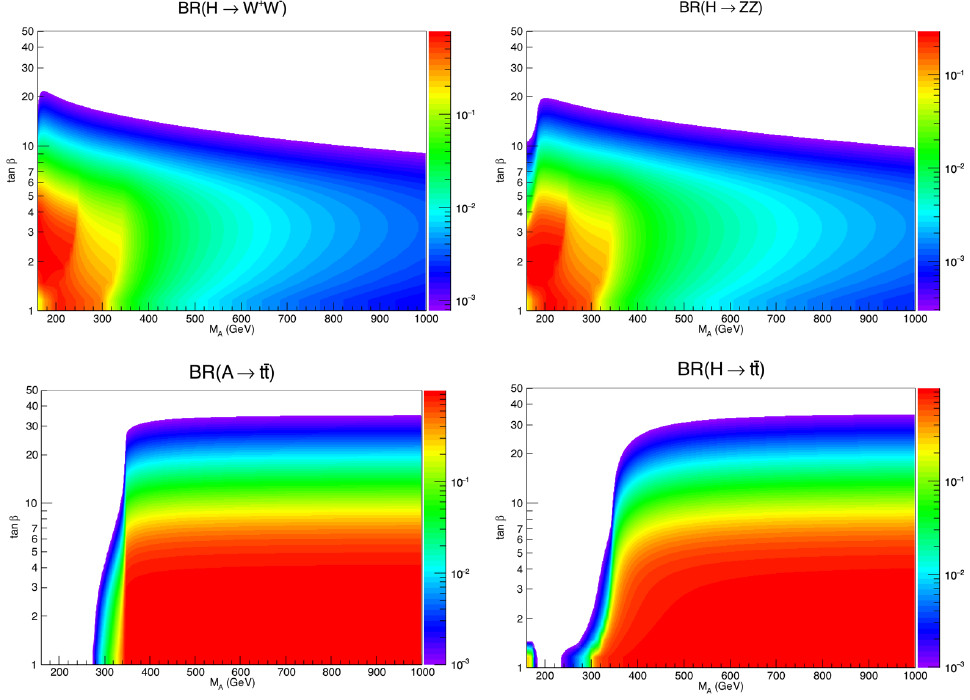}}\\
\emph{{Fig.3.}} {\emph{The branching ratios of the neutral 
Higgs bosons in the [tan$\beta$; M$_A$] parameter 
space of the hMSSM model, from \cite{12.}. }}\\
\ec

The production cross sections for 
A and H bosons are displayed in Fig. 4 
in the [tan$\beta$, M$_A$]
 hMSSM parameter space for 14 TeV at the LHC

\bec
{\includegraphics[width=0.85\textwidth]{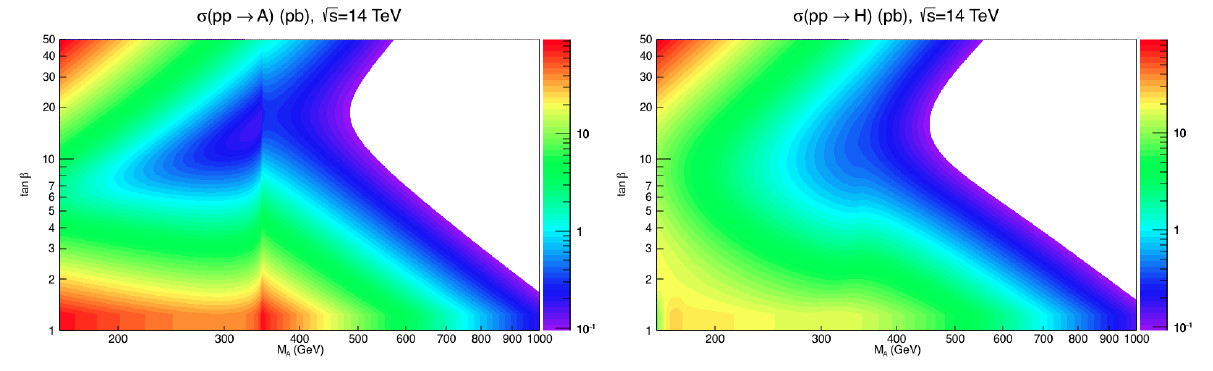}}\\
\emph{{Fig.4.}} {\emph{The production cross sections 
of the Higgs bosons A (left) 
and H (right) at the LHC with $\sqrt{s}$=14 TeV in the 
[tan$\beta$; M$_A$] hMSSM plane,
from \cite{12.}.}}\\
\ec

\section{Calculations of the production 
cross sections times branching fractions for Higgs bosons}

1) CP-even Higgs boson, H\\
Searches for heavy Higgs bosons by Run-2 ATLAS Collaboration 
at the LHC in the $H\rightarrow ZZ$ and $H\rightarrow WW$ decay channels 
are relevant due to the possibility of evidence for 
new particles beyond the Standard Model. The limits 
on $\sigma(pp\rightarrow H)\times BR(H\rightarrow ZZ)$  and 
$\sigma(pp\rightarrow H)\times BR(H\rightarrow WW)$ at 95$\%$ CL from \cite{13.} 
and \cite{14.} correspondingly are presented in Fig. 5
\bec
{\includegraphics[width=0.85\textwidth]{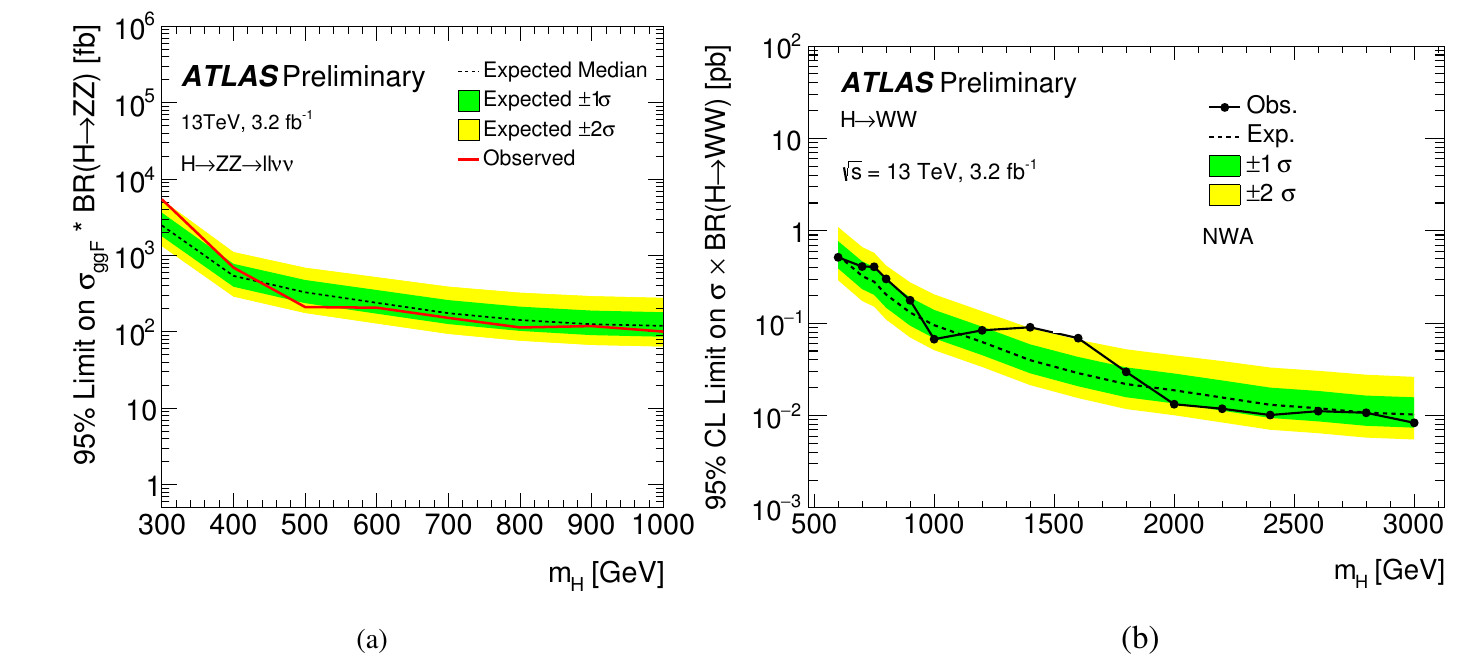}}\\
\emph{{Fig.5.}} {\emph{
Limits on $\sigma(pp\rightarrow H)\times BR(H\rightarrow ZZ)$ (a) and 
$\sigma(pp\rightarrow H)\times BR(H\rightarrow WW)$ (b) via
gluon-gluon fusion at 95$\%$ CL.}}\\
\ec
Using the restricted parameter set for [tan$\beta$; M$_A$] plane, 
presented in the previous section and computer programs SusHi 
\cite{15.}
and SOFTSUSY4.0 \cite{16.}, 
we calculated $\sigma(pp\rightarrow H)\times BR(H\rightarrow ZZ)$  and
$\sigma(pp\rightarrow H)\times BR(H\rightarrow WW)$ for $\sqrt{s}$=14 TeV 
at the LHC, presented in Fig. 6
\bec
{\includegraphics[width=0.81\textwidth]{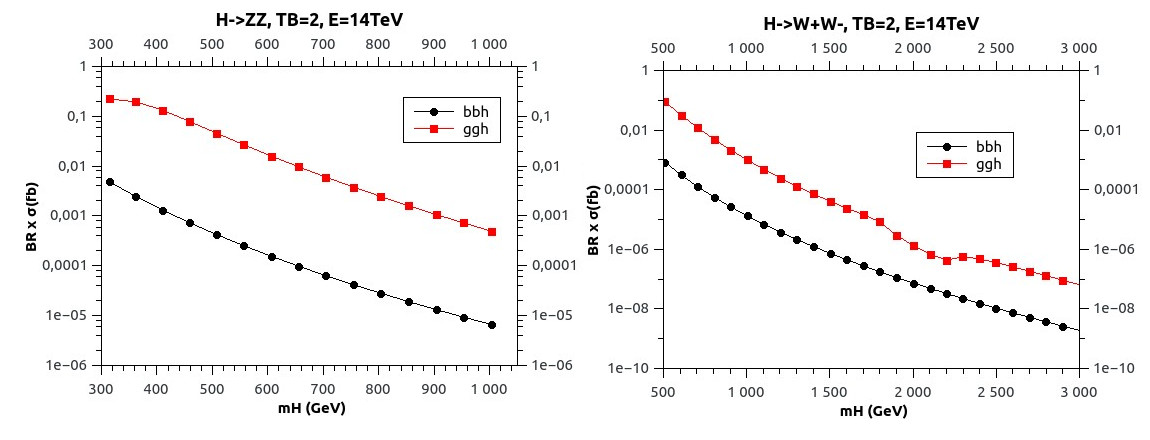}}\\
\emph{{Fig.6.}} {\emph{
$\sigma(pp\rightarrow H)\times BR(H\rightarrow ZZ)$ (left) and
$\sigma(pp\rightarrow H)\times BR(H\rightarrow WW)$ (right) 
for $\sqrt{s}$=14 TeV at the LHC. }}\\
\ec
From Fig. 6 we can see the increase in value $\sigma\times Br$  
for ggh fusion process compared with bbh fusion 
process of heavy Higgs boson, H production.
Since the branching ratios 
for the decays $H\rightarrow bb$ and $H\rightarrow tt$ 
are significant values according to our calculations with  
SOFTSUSY4.0 program, we have performed 
calculations of $\sigma(pp\rightarrow H)\times BR(H\rightarrow tt)$ 
and $\sigma(pp\rightarrow H)\times BR(H\rightarrow bb)$ for the planned 
at the LHC energy of 14 TeV, presented in Fig. 7

\bec
{\includegraphics[width=0.81\textwidth]{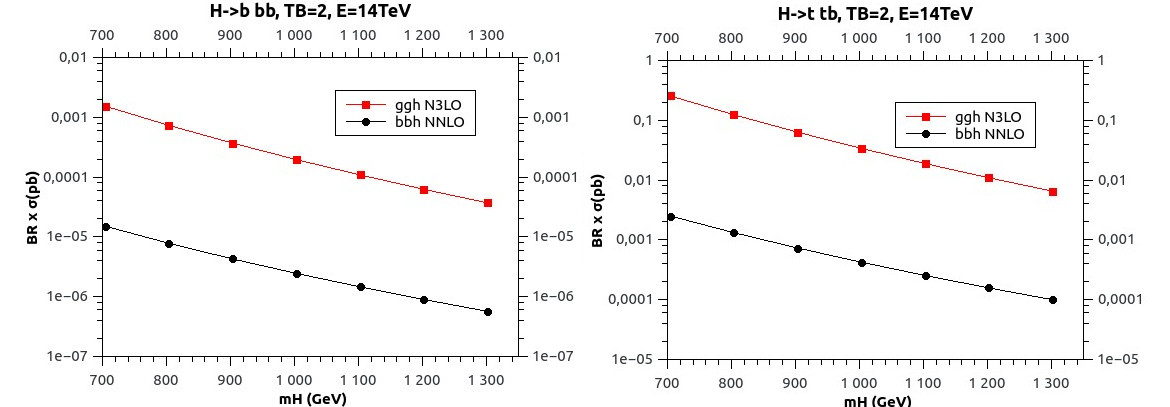}}\\
\emph{{Fig.7.}} {\emph{$\sigma(pp\rightarrow H)\times BR(H\rightarrow bb)$ 
(left) and $\sigma(pp\rightarrow H)\times BR(H\rightarrow tt)$ (right)
for $\sqrt{s}$=14 TeV at the LHC.}}\\
\ec

From the comparison of our calculations, presented 
above, we can see significant predominance of 
the values $\sigma\times Br$  for the second variant (Fig. 7) 
compared to the first one (Fig. 6). It is 
also important to stress the necessity of 
N3LO calculations for essential enlargement 
of the $\sigma\times Br$  value. \\ 
2) CP-odd Higgs boson, A\\
In this section we have considered the following 
decay processes of A boson: $A\rightarrow bb$ and $A \rightarrow tt$. 
The consideration of these processes of A boson 
decay is connected with the large value of 
branching ratio, that is represented in Fig.1. 
As we have calculated the process $ A\rightarrow Zh$ in 
\cite{17.}  and currently 
there are no other experimental data, for future 
experimental searches it was 
of interest to perform calculations for the two other decay 
channels from the three maximal. Using the 
computer programs SOFTSUSY4.0 and SusHi, we 
have performed the calculations of $\sigma\times Br$ for CP-odd 
Higgs boson, A. As the branching ratio for A 
boson is maximal for the decays $A\rightarrow bb$ and 
$A\rightarrow tt$ in the selected set of parameters, 
it was interesting to calculate $\sigma\times Br$  for this 
both processes over a wide range of boson 
masses, from 500 GeV to 3450 GeV. The 
results of our calculations are presented in Fig. 8
\bec
{\includegraphics[width=0.81\textwidth]{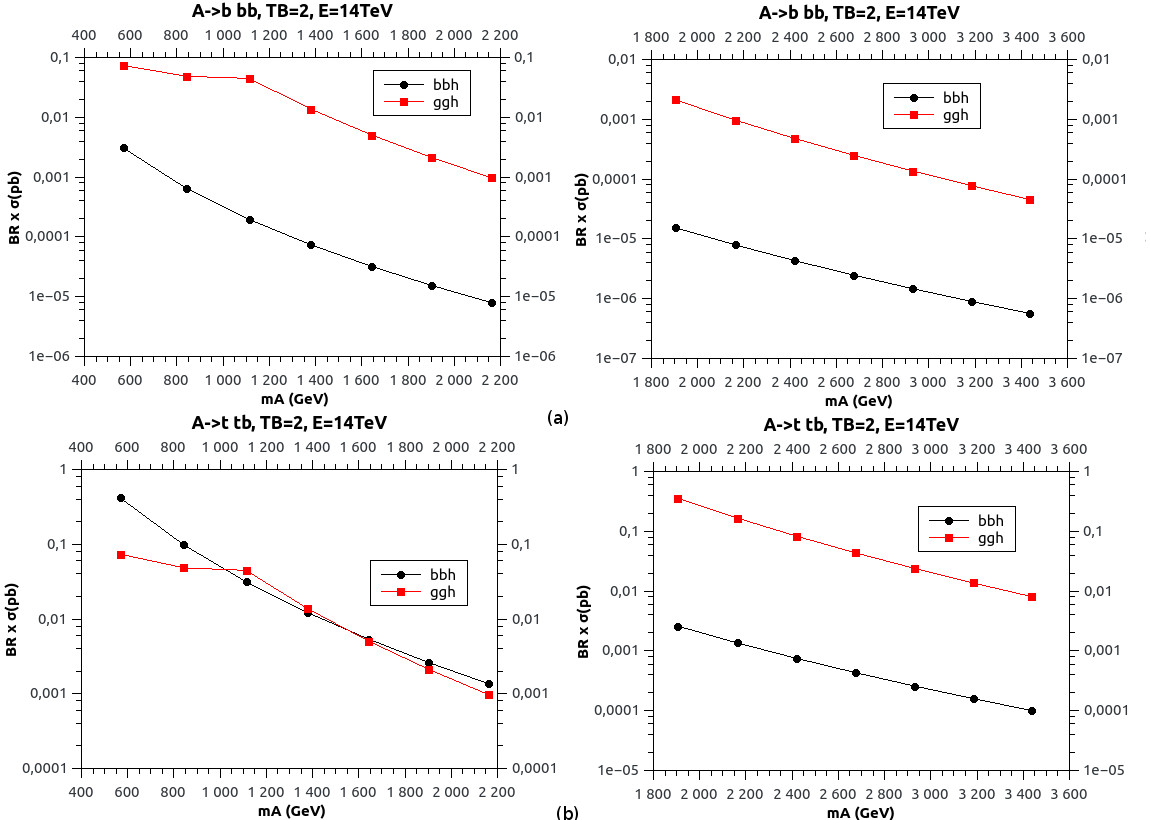}}\\
\emph{{Fig.8.}} {\emph{
$\sigma(pp\rightarrow A)\times BR(A\rightarrow bb)$ in the mass
range 500-2200 GeV (left) and 1800-3450 GeV (right) (a)
and $\sigma(pp\rightarrow A)\times BR(A\rightarrow tt)$ 
in the mass
range 500-2200 GeV (left) and 1800-3450 GeV (right) (b).
}}\\
\ec

From Fig. 8 we can see the predominance of 
the ggh process of A boson formation over 
the bbh one except for the (b) case of 
$A\rightarrow tt$ process in the mass range of 500-2200 GeV 
with interesting intersection points between 
bbh and ggh processes. It is also necessary 
to stress the largest value of $\sigma\times Br$ for the  
smallest masses, $m_A$, what is easily explained 
in connection with the lower mass of the Higgs boson A. \\
3) charged Higgs bosons, H$\pm$\\
As is known \cite{18.}, 
the production of charged Higgs boson 
depends on its mass and for m$_{H^{+}}$ $>$ m$_t$, 
H$^+$ production mode is associated 
with a top quark, as illustrated in Fig. 9
\bec
{\includegraphics[width=0.31\textwidth]{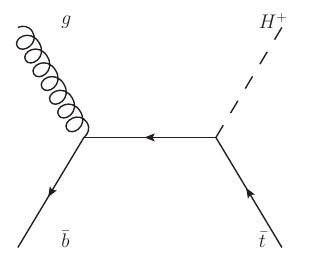}}\\
\emph{{Fig.9.}} {\emph{
Leading-order Feynman diagram for the production of H$^+$ 
in association with a top quark in five flavor scheme.}}\\
\ec
In Fig. 10 are shown the expected and 
observed limits for the production of 
$H^+\rightarrow tb$ in association with a top quark, 
bands for 68$\%$ (in green) and 95$\%$ (in yellow) 
confidence intervals and the signal 
prediction in the m$^{mod-}_h$  benchmark scenario 
of the MSSM \cite{19.}. 
\bec
{\includegraphics[width=0.55\textwidth]{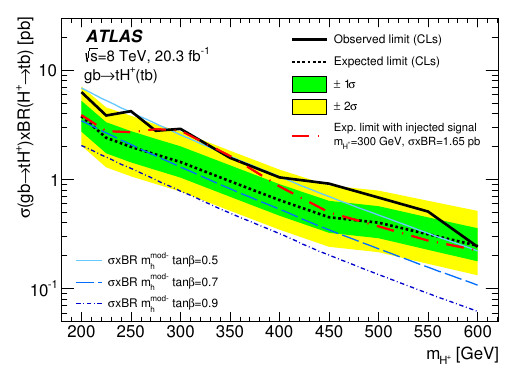}}\\
\emph{{Fig.10.}} {\emph{
Expected and observed limits for the production o
f H$^+\rightarrow tb$ 
in association with a top quark, from \cite{18.}}.}\\
\ec
As model points with $0.5\leq$tan$\beta\leq 0.6$, tan$\beta\approx 0.5$, 
tan$\beta$=0.7 and tan$\beta$=0.9  are excluded in the 
H$^+$ mass range of 200-600 GeV obtained also in other 
scenarios of MSSM, it would be interesting to 
do the calculations of $\sigma\times Br$ for tan$\beta$=2. For the 
studying of properties of charged Higgs bosons, 
H$^{\pm}$, we have used the set of parameters of MSSM 
model to calculate the cross-sections of tH$^+$ 
production with the help of the software program 
PROSPINO \cite{20.}
with data implemented from the 
latest computer program SOFTSUSY4.0. 
The corresponding results for 
$\sigma(pp\rightarrow tH^+)BR(H^+\rightarrow tb)$, obtained 
for the parameter set of tan$\beta$=2 and for the 
energy of 14 TeV in the mass range of  m$H^+$=500-1200 GeV  
are presented in Fig.11
\bec
{\includegraphics[width=0.55\textwidth]{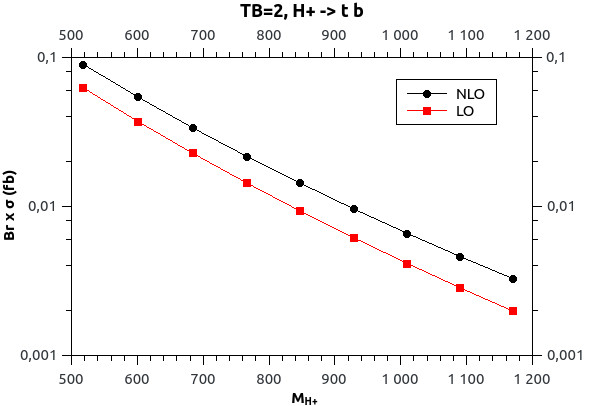}}\\
\emph{{Fig.11.}} {\emph{
$\sigma(pp\rightarrow tH^+)BR(H^+\rightarrow tb)$
for 14 TeV at the LHC in the mass range of m$_{H^+}$=500-1200 GeV.}}\\
\ec
Another most visible decay channel of a 
charged Higgs boson is $H^+\rightarrow\tau\nu$. Its searches in 
association with a single top quark were 
performed by ATLAS Collaboration at the 
LHC with proton--proton collision at $\sqrt{s}$=13 TeV 
corresponding to an integrated luminosity 
of 3.2 fb$^{-1}$. The analysis of experimental 
data leads to 95$\%$ CL upper limits on 
the $\sigma(pp\rightarrow [b]tH^{\pm})BR(H^{\pm}\rightarrow\tau\nu)$, 
between 1.9 pb and 15 fb, for m$_{H^+}$=200-2000 GeV, 
that is presented in Fig. 12.
\bec
{\includegraphics[width=0.61\textwidth]{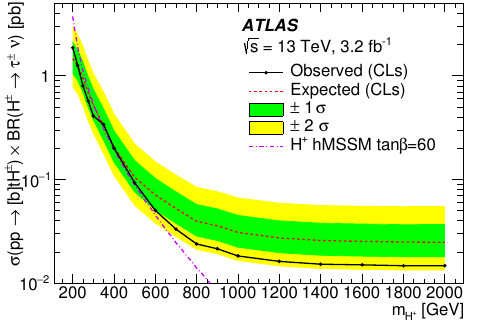}}\\
\emph{{Fig.12.}} {\emph{
Observed and expected 95$\%$ CL exclusion limits for heavy 
charged Higgs boson production as a function 
of m$_{H^+}$, from \cite{21.}.}}\\
\ec
From these experimental data tan$\beta$ = 42--60 
for m$_{H^+}$=200 GeV and tan$\beta$=60 for the H$^+$ mass 
range from 200 to 340 GeV were excluded. So we 
have considered two cases of tan$\beta$=2 and 30 
for comparison of the value of  $\sigma(pp\rightarrow [b]tH^{\pm})
BR(H^{\pm}\rightarrow\tau\nu)$ 
for these two cases, presented in Fig. 13. 
\bec
{\includegraphics[width=0.84\textwidth]{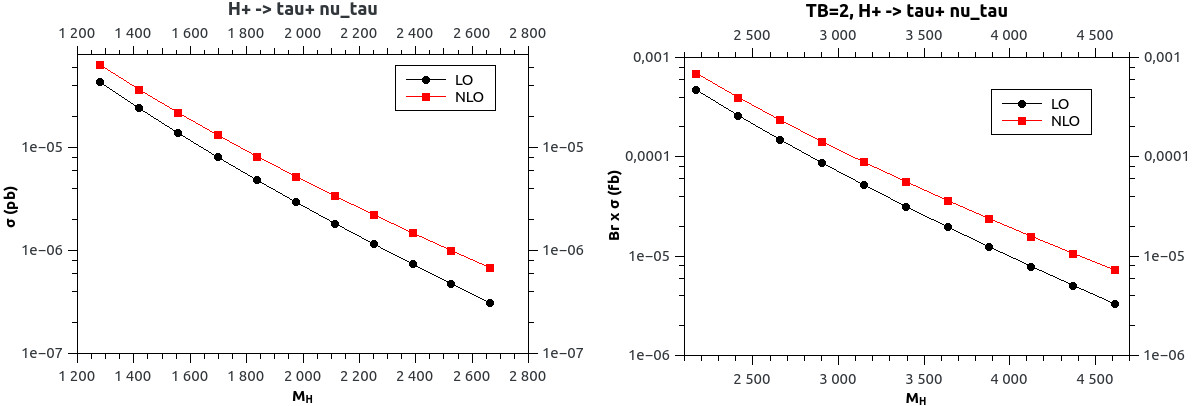}}\\
\emph{{Fig.13.}} {\emph{
$\sigma(pp\rightarrow [b]tH^{\pm})
BR(H^{\pm}\rightarrow\tau\nu)$ for (a) tan$\beta$=30 
in the mass range m$_H^+$= 1200-2650 GeV (b) and
tan$\beta$=2 in the mass range m$_H^+$=2200-4600 GeV
with the planned 14 TeV at the LHC.}}\\
\ec
From Fig. 13 the predominance in the value of 
$\sigma(pp\rightarrow [b]tH^{\pm})
BR(H^{\pm}\rightarrow\tau\nu)$  for the variant (a) 
is obvious but we can see the larger values of 
 $\sigma(pp\rightarrow [b]tH^{\pm})
BR(H^{\pm}\rightarrow\tau\nu)$ for tan$\beta$=30  in 
the range of the mass intersection of charged 
Higgs boson, m$_H^+$=2200-2650 GeV for (a) and (b) variants.
In addition, it is known that for $m_{H^+}>m_t$  the dominant 
decay of H$^+$ is $H^+\rightarrow tb$, but for large values of tan$\beta$ 
is observed a substantial contribution from $H^+\rightarrow\tau\nu$  
\cite{21.}. For comparison 
we calculated $\sigma(pp\rightarrow tH^+)BR(H^+\rightarrow tb)$ for 
tan$\beta$=30 for 14 TeV at the LHC, presented in Fig.14
\bec
{\includegraphics[width=0.61\textwidth]{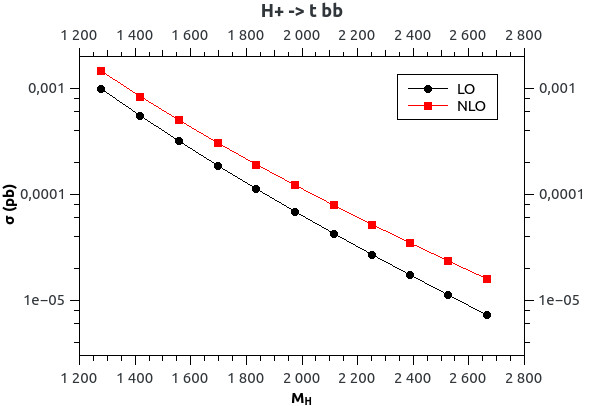}}\\
\emph{{Fig.14.}} {\emph{
$\sigma(pp\rightarrow tH^+)BR(H^+\rightarrow tb)$ for 14 TeV at the LHC
in the mass range of m$_H^+$=1200-2650 GeV.}}\\
\ec
From the Fig. 14 and 13 (a) it can be concluded 
about the largest values of $\sigma(pp\rightarrow tH^+)BR(H^+\rightarrow tb)$
 in contrast 
with $\sigma(pp\rightarrow [b]tH^{\pm})BR(H^{\pm}\rightarrow\tau\nu)$ 
for the same tan$\beta$=30, but the increase of the value 
$\sigma(pp\rightarrow [b]tH^{\pm})BR(H^{\pm}\rightarrow\tau\nu)$ for the larger 
tan$\beta$ was stressed above.

\section{Conclusion}

Using the restricted parameter set of the hMSSM model, 
presented in \cite{5.} and \cite{12.} for 
the extended sector of Higgs bosons 
as well as the latest experimental data on the observed and 
expected CL exclusion limits for Higgs boson production, 
performed by ATLAS Collaboration \cite{13.}, 
\cite{14.}, \cite{18.}, \cite{21.} with the help of software programs 
SOFTSUSY4.0, SusHi and PROSPINO we have calculated $\sigma\times Br$ 
for CP-even Higgs boson, H, CP-odd Higgs boson, 
A and charged Higgs bosons, H$^{\pm}$. From our calculations 
we can conclude about the large values of the  $\sigma\times Br$  at 
small tan$\beta$=2 for chosen decay channels of Higgs bosons for the
energy at the LHC of 14 TeV. But for 
the charged Higgs boson are obtained another results, 
that are connected with larger values of tan$\beta$.

\label{page-last} 
\label{last-page}
\end{document}